\documentclass{article}
\usepackage{hiph-preprint}
\usepackage{graphicx}
\usepackage{amsmath}
\volnumber{22} \issuenumber{1} \edyear{2005}                             %|
\frompage{000} \topage{000}                                              %|
\recrevdate{28 October 2005}                                              %|
\def\rightmark{Numerical simulation of non-extensive Boltzmann equation}
\def\leftmark{T.S. Bir\'o and G. Purcsel}
 
\title{Numerical simulation of non-extensive\\ Boltzmann equation} 
\authors{ 
{Tam\'as S. Bir\'o$^{1,2}$ and G\'abor Purcsel$^{1,2}$ %
\index{Bir\'o, T.S.} % Abbreviated names of the author(s),
\index{Purcsel, G.} % to be inserted for use in the volume index
}\\[2.812mm]
{\normalsize
\hspace*{-8pt}$^1$ Institut f\"ur Theoretische Physik, Justus-Liebig-Universit\"at,\\
					Gie{\ss}en, Germany,\\ [0.2ex] 
\hspace*{-8pt}$^2$ KFKI Research Institute for Particle and Nuclear Physics,\\
					Budapest, Hungary\\
}}
 
\abstract{We present first results of the development of a test particle simulation for solving non-extensive extensions of the elastic two-particle Boltzmann equation. Stationary one-particle energy distributions with power-law tail are obtained.}


\makeindex
\begin{document}
 
\maketitle

\section{Introduction}\label{intro}
There are several experimental evidences on power-law tailed statistical distributions of single particle energy. One can find power-law tailed distributions also in a wide range of other fields of science \cite{c1,c2,c3}.
 Our aim here is to reproduce exponentially starting and power-law tailed spectra in the framework of the non-extensive thermodynamics, instead of fitting the low- and high-energy parts of the spectra separately. To achieve our goal we use a generalized Boltzmann equation \cite{TSBPRL}, which accomodates a general, associative energy composition rule, defining this way a certain realization of non-extensive thermodynamics. Another way to generalize the Boltzmann equation \cite{GK} preceeded our approach, it can be connected to our extension of the Boltzmann theory by requiring the same stationary solution \cite{TSBGK}. To each composition rule a strict monotonic function can be constructed which transforms it to a normal extensive quantity. In the theory of formal groups this function is called the logarithm of the group. We investigate also the evolution of the one-particle distribution in the framework of this two-body Boltzmann equation supported by a non-extensive energy composition rule.

%%%%%%%%%%%%%%%%%%%%%%%%%%%%%%%%%%%%%%%%%%%%%%%%%%%%%%%%%%%%%%%%%%%%%%%%%%%%%%
%%%%%%%%%%%%%%%%%%%%%%%%%%%%%%%%%%%%%%%%%%%%%%%%%%%%%%%%%%%%%%%%%%%%%%%%%%%%%%
%%%%%%%%%%%%%%%%%%%%%%%%%%%%%%%%%%%%%%%%%%%%%%%%%%%%%%%%%%%%%%%%%%%%%%%%%%%%%%

\section{Non-Extensive Boltzmann Equation}
	In the non-extensive extension of the Boltzmann equation we keep the original factorizing form for the two-body distributions,
	\begin{equation}
		\frac{\partial}{\partial t} f_1 = \int_{234} w_{1234} \left( f_3f_4-f_1f_2 \right),
	\end{equation}
	but in the transition rate
	\begin{equation}
		w_{1234} = M_{1234}^2\cdot \delta\left(\left(\vec p_1+\vec p_2\right)-\left(\vec p_3+\vec p_4\right)\right) \cdot \delta\left(h\left(E_1,E_2\right)-h\left(E_3,E_4\right)\right),
	\end{equation}
	the two-body energy composition is not necessarily extensive, (\(E_{12} = h\left(E_1,E_2\right) \not= E_1+E_2\)).
	
	Clearly \(E_i=\sqrt{\vec p_i^2+m_i^2}\) are regarded as asymptotic, purely kinetic energies, while \(h(E_1,E_2)\) may contain further contributions stemming from the pair interaction. It is not trivial whether these can always be divided to one-particle contributions, supporting a quasi-particle picture. In the followings we demonstrate that under quite general assumptions about the value of the function \(h(x,y)\) it can, however, be done.
	
	We assume that the generalized energy sum is associative
	\begin{equation}
			h\left(h\left(x,y\right),z\right) = h\left(x,h\left(y,z\right)\right);
	\end{equation}
	a reasonable assumption for approaching the thermodynamical limit. Due to a mathematical theorem one finds a function \(X(h)\), which maps the energy composition rule to additivity, i.\ e.
	\begin{equation}
			%X\left(h\left(x,y\right)\right) = X\left(x\right) + X\left(y\right)
			X\left(h\right) = X\left(x\right) + X\left(y\right) \label{Xadd}
	\end{equation}
	is fulfilled. This solution is unique up to a constant multiplicative factor. %, \(Y\left(h\right) = c\cdot X\left(h\right)\).
	In this case the stationary solution of the non-extensive Boltzmann equation is given by
	\begin{equation}
			%f\left(E\right) \sim e^{-X\left(E\right)/T}
			f\left(\vec p\right) = \frac1Z\: e^{-X\left(E\right)/T}.
	\end{equation}
	%Here \(X(E)\) can be considered as a statistical quasi-energy while \(E(\vec p)\) is the free dispersion relation.

%\section{Composition rule examples}
Table \ref{comprules} shows a couple of interesting composition rules, which lead to widely studied statistics.
\begin{table}[h]
	\begin{center}
	\begin{tabular}{||c|c|c|c||}
		\hline
		\hline
		& \(h(x,y)\) & \(X(E)\) & \(Z\cdot f(E)\) \\
		\hline
		&&&\\
		Gibbs & \(x+y\) & \(E\) & \(e^{-E/T}\) \\
		&&&\\
		Tsallis & \(x+y+axy\) & \(\frac1a\ln(1+aE)\) & \((1+aE)^{-1/(aT)}\) \\
		&&&\\
		L\'evy & \((x^b+y^b)^{1/b}\) & \(\frac1a(aE)^b\) & \(e^{-(aE)^b/(aT)}\) \\
		&&&\\
		R\'enyi & \(axy\) & \(\frac1a\ln(aE)\) & \((aE)^{-1/(aT)}\) \\
		&&&\\
		\hline
		\hline
	\end{tabular}
	\end{center}
%	\vspace*{-1.0cm}
	\caption{Examples for the \(h(x,y)\) composition rule.}
	\label{comprules}
\end{table}

%%%%%%%%%%%%%%%%%%%%%%%%%%%%%%%%%%%%%%%%%%%%%%%%%%%%%%%%%%%%%%%%%%%%%%%%%%%%%%
%%%%%%%%%%%%%%%%%%%%%%%%%%%%%%%%%%%%%%%%%%%%%%%%%%%%%%%%%%%%%%%%%%%%%%%%%%%%%%
%%%%%%%%%%%%%%%%%%%%%%%%%%%%%%%%%%%%%%%%%%%%%%%%%%%%%%%%%%%%%%%%%%%%%%%%%%%%%%

\section{Numerical simulation}
We solve numerically the non-extensive Boltzmann equation by a parton cascade simulation. We use several different initial momentum distributions, and then make random binary collisions between randomly chosen pairs of particles. By doing so we apply the following rules:
	\begin{equation}
		X\left(E_1\right) + X\left(E_2\right) = X\left(E_3\right) + X\left(E_4\right)  \label{bcr1}
	\end{equation}
	\begin{equation}
		\vec p_1 + \vec p_2 = \vec p_3 +\vec p_4  \label{bcr2}
	\end{equation}
After selecting two colliding particles in the simulation, we find the value for the new momentum of the first particle (\(\vec p_3\)) satisfying the above constraints but otherwise random. Then applying the rules in eqs.\ (\ref{bcr1}) and (\ref{bcr2}) we can calculate the momentum of the second outgoing particle (\(\vec p_4\)). Presently we use the free dispersion relation for massless particles (\(E_i(\vec p_i) = \left|\vec p_i\right|\)). 
A typical simulation includes \(10^6-10^7\) collisions. After \(3-5\) collisions per particle on the average, the one-particle distribution reaches its stationary form.

There are several conserved quantities in the simulation, 
	\begin{equation}
		X\left(E_{tot}\right) = \sum_{i=1}^N X\left(E_i\right), \;\;\;\;
		\vec P = \sum_{i=1}^N \vec p_i, \;\;\;\;
		N = \sum_{i=1}^N 1,
	\end{equation}
which are monitored during the simulation.

%%%%%%%%%%%%%%%%%%%%%%%%%%%%%%%%%%%%%%%%%%%%%%%%%%%%%%%%%%%%%%%%%%%%%%%%%%%%%%
%%%%%%%%%%%%%%%%%%%%%%%%%%%%%%%%%%%%%%%%%%%%%%%%%%%%%%%%%%%%%%%%%%%%%%%%%%%%%%
%%%%%%%%%%%%%%%%%%%%%%%%%%%%%%%%%%%%%%%%%%%%%%%%%%%%%%%%%%%%%%%%%%%%%%%%%%%%%%

\section{Kinematics with non-extensive rules}
Most power-law tailed spectra emerging from the Tsallis thermodynamics \cite{TS} are based on the following composition rule
	\begin{equation}
		h(x,y) = x+y+axy.
		\label{TsRule}
	\end{equation}
It can be regarded as the sub-leading order expansion of eq.\ (\ref{Xadd}) for small arguments. The traditional Boltzmann distribution is recovered in the case of \(a=0\).

In one certain binary collision the total quasi-energy of the two particles is conserved, so after choosing  randomly  a new energy value for the first outgoing particle (\(E_3\)), we can calculate the new energy value of the second outgoing particle (\(E_4\)), by inverting eq.\ (\ref{TsRule}),
	\begin{equation}
		E_4 = \frac{E_{12} - E_3}{1 + a E_3},  \;\;\;\;\;\;  E_{12} = E_1+E_2+aE_1E_2.
		\label{TsRuleInv}
	\end{equation}

	\begin{figure}[!h]
	\begin{center}	 
	\resizebox{7.8 cm}{!}{\includegraphics*{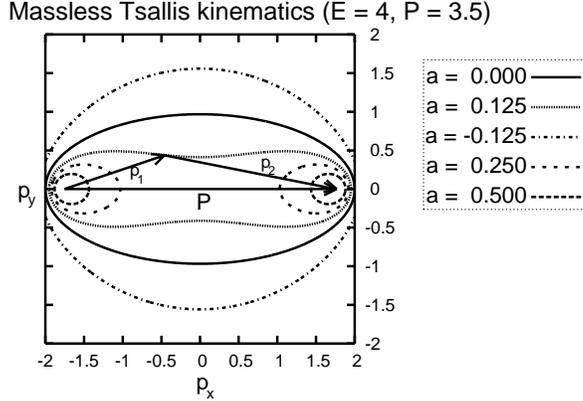}}
%	\vspace*{-0.7cm}
	\caption{Kinematics with \(h(x,y)=x+y+axy\) rule for \(E = 4\) and \(P = 3.5\).}
	\label{TsKin2}
	\end{center}	 
	\end{figure}
On Fig.\ \ref{TsKin2} an example is shown, how one can obtain \(\vec p_4\) and \(\vec p_3\) when the energy and the momenta of the incoming two particles are known.
This figure shows cross sections of the surfaces in the phase space, whereon the endpoint of \(\vec p_1\), which is also the startpoint of \(\vec p_2\), lies on. \(\vec p_3+\vec p_4=\vec p_1+\vec p_2\ = \vec P\), and the new momentum \(\vec p_3\) should also point to this surface. For the composition rule in eq.\ (\ref{TsRule}) the endpoints of the vector \(\vec p_3\) constitute a surface of revolution of fourth-order curves. The special case of an ellipse emerges for \(a=0\).

%%%%%%%%%%%%%%%%%%%%%%%%%%%%%%%%%%%%%%%%%%%%%%%%%%%%%%%%%%%%%%%%%%%%%%%%%%%%%%
%%%%%%%%%%%%%%%%%%%%%%%%%%%%%%%%%%%%%%%%%%%%%%%%%%%%%%%%%%%%%%%%%%%%%%%%%%%%%%
%%%%%%%%%%%%%%%%%%%%%%%%%%%%%%%%%%%%%%%%%%%%%%%%%%%%%%%%%%%%%%%%%%%%%%%%%%%%%%

\section{Spectra with non-extensive rules}
\begin{figure}[]
	\begin{center}	
		\resizebox{10.8 cm}{!}{\includegraphics*{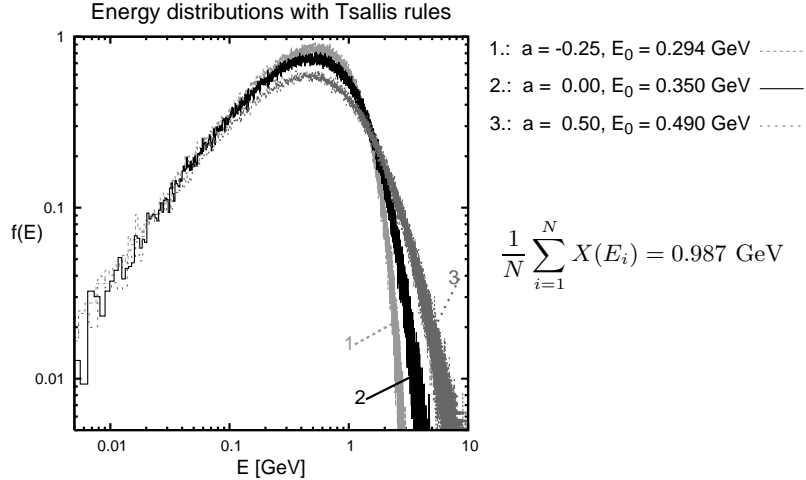}}
%	\vspace*{-0.7cm}
		\caption{One-particle energy distributions with \(h(x,y)=x+y+axy\) rule.}
		{\small
		\vspace{-5 cm}
		\[ \hspace{6 cm}  \frac1N\sum_{i=1}^N X(E_i) = 0.987 \text{ GeV} \]
		\vspace{3 cm}
		}
		\label{Spectra}
	\end{center}	
	\end{figure}

The result of three simulations with the energy composition rule of eq.\ (\ref{TsRule}) is shown on Fig.\ \ref{Spectra}. In all of the three cases the number of particles (\(10^5\)) and the number of collisions (\(10^6\)) are the same, the particles are massless. \(E_0\) is chosen to get the same value for \(X(E_{tot})\) for each value of the parameter \(a\).

The \(a=0\) case leads to the traditional Boltzmann distribution (curve 2), while the one with positive Tsallis parameter (\(a=0.5\)) has a power-law tail above the exponential curve  (curve 3) and the distribution with negative parameter (\(a=-0.25\)) goes below the exponential (curve 1).

%%%%%%%%%%%%%%%%%%%%%%%%%%%%%%%%%%%%%%%%%%%%%%%%%%%%%%%%%%%%%%%%%%%%%%%%%%%%%%
%%%%%%%%%%%%%%%%%%%%%%%%%%%%%%%%%%%%%%%%%%%%%%%%%%%%%%%%%%%%%%%%%%%%%%%%%%%%%%
%%%%%%%%%%%%%%%%%%%%%%%%%%%%%%%%%%%%%%%%%%%%%%%%%%%%%%%%%%%%%%%%%%%%%%%%%%%%%%

\section{Conclusions}
We demonstrated how to simulate power-law tailed stationary distributions in a parton cascade code. As a generic example the cut power-law distribution was generated. We have found a unique non-exponential stationary distribution for each non-extensive energy addition rule. The non-extensivity parameter is yet to be derived from microscopic interactions. We conclude that power-law tails in one-particle spectra are compatible with stationary states of generalized non-extensive kinetic models.

\section*{Acknowledgments}
Discussions with Dr.\ G\'eza Gy\"orgyi at E\"otv\"os University and Dr.\ Antal Jakov\'ac at the Budapest University of Technology and Economics are hereby gratefully acknowledged. This work has been supported by the Hungarian National Science Fund OTKA (T049466) and the Deutsche Forschungsgemeinschaft.

\vfill\eject
\end{document}